\documentclass[12pt]{article}
\usepackage{amsmath,amsfonts,amsthm,graphicx}
\usepackage[authoryear,comma]{natbib}
\usepackage{algorithm}
\usepackage{algpseudocode,geometry}

\usepackage{hyperref}
\hypersetup{citecolor=blue,colorlinks=true}

\usepackage{autonum}

\theoremstyle{plain}
\newtheorem{theorem}{Theorem}
\newtheorem{lemma}{Lemma}

\theoremstyle{definition}

\theoremstyle{remark}

\newcommand{\x}{{\boldsymbol{x}}}
\newcommand{\X}{{\boldsymbol{X}}}
\newcommand{\bsigma}{{\boldsymbol{\Sigma}}}
\newcommand{\s}{{\boldsymbol{S}_{\A n}}}
\newcommand{\Z}{{\boldsymbol{Z}}}
\newcommand{\y}{{\boldsymbol{y}}}
\newcommand{\bu}{{\boldsymbol{u}}}
\newcommand{\A}{{\mathcal{A}}}
\newcommand{\bbeta}{{\boldsymbol{\beta}}}
\newcommand{\btheta}{{\boldsymbol{\theta}}}

\DeclareMathOperator{\tr}{tr}
\newcommand{\be}{\begin{equation}}
\newcommand{\ee}{\end{equation}}
\newcommand{\ba}{\begin{eqnarray}}
\newcommand{\ea}{\end{eqnarray}}
\newcommand{\bee}{\begin{equation*}}
\newcommand{\eee}{\end{equation*}}
\newcommand{\baa}{\begin{eqnarray*}}
	\newcommand{\eaa}{\end{eqnarray*}}
\newcommand{\nn}{\nonumber\\}
\newcommand{\bo}[1]{\mathbf{#1}}

	\title{Robust Variable Selection Criteria for the Penalized Regression}
\author{Abhijit Mandal\\
Department of Mathematics, Wayne State University\\ 
and\\
Samiran Ghosh\\ 
Department of Family Medicine \& Public Health Science and \\
Centre of Molecular Medicine and Genetics, Wayne State University
}

\begin{document}
	
	\maketitle

	\begin{abstract} 
		
		We propose a robust variable selection procedure using a divergence based M-estimator combined with a penalty function. It produces robust estimates of the regression parameters and simultaneously selects the important explanatory variables. An efficient algorithm based on the quadratic approximation of the estimating equation is constructed. The asymptotic distribution and the influence function of the regression coefficients are derived. The widely used model selection procedures based on the Mallows's $C_p$ statistic and Akaike information criterion (AIC) often show very poor performance in the presence of heavy-tailed error or outliers. For this purpose, we introduce  robust versions of these information criteria based on our proposed method. The simulation studies show that the robust variable selection technique outperforms the classical likelihood-based techniques in the presence of outliers. The performance of the proposed method is also explored through the real data analysis.

\end{abstract}
	
	\noindent{\textbf{MSC2010 subject classifications}}: 62J07, 62F35.
	

	\section{Introduction}

	We address the development of a robust method for modeling and analyzing high-dimensional data in the presence of outliers. Due to advanced technology and wide source of data collection, a high-dimensional data is available in several fields including healthcare, bioinformatics, medicine, epidemiology, economics, finance, sociology and climatology. In those data-sets, outliers are commonly encountered generally due to heterogeneous sources or effect of some confounding variables. The standard approaches often fail to model such data and produce misleading information. The modeling approaches can also be challenged by model misspecification and  heavy-tailed error distribution. Thus, a suitable robust statistical method is essential to analyze these data which can properly eliminate the effect of outliers. 
	
	In the initial stage of modeling, generally, a large number of predictors are included to get maximum information from data. However, in practice, very few predictors contain relevant information about the response variable. Thus, variable selection is an important topic in regression analysis when there are large number of predictors. It enhances the predictability power of the model, and reduces the chance of over-fitting. It also provides a better understanding of the underlying process that generated the data, and gives a faster and more cost-effective predictors. Including many predictors in the final model unnecessarily adds noise to the estimation of main quantities that we are interested in. The classical regression analysis is badly affected by multi-collinearity when too many variables tries to do the same job in explaining the response variables. Therefore, to explore the data in the simplest way, one needs to remove redundant predictors. 
 As a powerful tool for selecting the subset of important predictors
	associated with responses, penalization plays a significant  role in the
	high-dimensional statistical modeling.  Methods that have been proposed include the
	bridge estimator \citep{frank1993statistical},
	least absolute shrinkage and selection operator or LASSO \citep{MR1379242}, the
	smoothly clipped absolute deviation or SCAD \citep{MR1946581}, the elastic net \citep{MR2137327}, the adaptive LASSO \citep{MR2279469} and the minimum concave penalty approach or MCP \citep{MR2604701}. 
The statistical properties of these methods are extensively studied in the literature, however,
most of these existing methods such as penalized least-squares or penalized likelihood \citep{fan2011nonconcave} are designed for light-tailed distributions. 
	Not only these methods  break down in the presence of outliers, but also the effect of outliers is not  well studied for many variable selection techniques (\citealp{MR2836768}).  Therefore, the robust variable selection, that can withstand the effect of outliers, is essential to model and analyze such data.

	In the literature, robust regularization methods such as the least absolute deviation (LAD) regression and quantile regression have been widely used for variable selection  \citep{MR2424800, MR2418651}. 
	 \cite{MR2797841, MR2949353}  studied the penalized quantile regression in high-dimensional sparse models where the dimensionality could be larger than the sample size.   \cite{MR3025129}  obtained bounds on the prediction error of a large class of $L_1$-penalized estimators, including quantile regression.  \cite{MR3189488, MR2815779} introduced the penalized quantile regression with the weighted $L_1$-penalty for robust regularization. Variable selection methods based on  M-estimators are addressed in (\citealp{MR2836768, MR2796868,kawashima2017robust}). In this paper, we propose a variable selection method based on the density power divergence (DPD) measure \citep{MR1665873}. 

	A penalized variable selection method uses a regularized parameter in the penalty function which controls the complexity of the model. 
	A commonly used method is the cross validation technique where the model parameters are estimated from the training data, and then the regularized parameter is selected from the remaining test data \citep{golub1979generalized}. However, if there are outliers in data, both the estimation and testing process may be severely affected. Therefore, the classical cross validation technique may not work properly in the presence of outliers. Moreover, the cross validation technique is computationally intensive. For the same reason, the bootstrap based methods may also fail in the presence of outliers. Another widely used technique is the information based criteria for the model selection. The Mallows's $C_p$ statistic \citep{mallows1973some},  the Akaike information criterion (AIC) \citep{MR0483125} and the Bayes information criterion (BIC) \citep{MR0468014} play an important role in high-dimensional data analysis.
	Unfortunately, as most selection criteria  are developed
	based on the ordinary least-squares (OLS) estimates, their  performance
	under heavy-tailed errors is very poor. \cite{MR784761, MR1294082} modified the classical selection criteria using the Huber's M-estimator. Consequently, \cite{MR1045193}  derived a set of useful model selection criteria based on the LAD estimates. Despite their usefulness, these LAD-based variable selection criteria,
	have some limitations -- the major one being  the computational
	burden \citep{MR2380753}. 
	To address the deficiencies of traditional model selection
	methods, we propose two information criteria using robust estimators based on the density power divergence. The detailed theoretical derivations are provided for these methods, and their performance is verified from the simulation studies and a real data example. 
	
	
The rest of the paper is organized as follows. Section \ref{sec:intro} gives the background of the classical penalized regression analysis. Our proposed method for the robust penalized regression is introduced in Section \ref{sec:robust_reg}.  In Sections \ref{sec:algo} and \ref{sec:asymp}, we presented the computation algorithm and the asymptotic distribution, respectively,  of the proposed estimator. The robustness properties of the estimator is discussed from the view of the influence function analysis in Section \ref{sec:inf}. Then, in Section  \ref{sec:selection}, two information criteria for model selection are proposed as robust versions of the Mallow's $C_p$ statistic and Akaike information criterion (AIC). An extensive simulation study and a real data analysis are presented to explore the effectiveness of the proposed method in Sections \ref{sec:simulatoin} and \ref{sec:data}, respectively. Some concluding remarks are given in Section \ref{sec:conc}, and the proofs and theoretical derivations are provided in the supplementary materials. 
	
	\section{Classical Penalized Regression} \label{sec:intro}
	Suppose the pair $(y_i, \x_i)$ denote the observation from the $i$-th subject, where   $y_i \in \mathbb{R}$ is the response variable and  $\x_i \in \mathbb{R}^{p+1}$ is the set of linearly independent predictors with the first element of $\x_i$ being one for the intercept parameter. 
	Consider the following linear regression model:
	\be
	y_i = \x_i^T \bbeta + \epsilon_i, \ \ \ i = 1, 2, \cdots, n,
	\label{reg_model}
	\ee
	where  $\bbeta = (\beta_0, \beta_1, \cdots, \beta_p)^T$ is the regression coefficient, and $\epsilon_i$ is the random error. We assume that the error term $\epsilon_i  \overset{iid}\sim N(0, \sigma^2)$.     So, we have
	$
	y_i \sim N(\x_i^T \bbeta, \sigma^2), \  i = 1, 2, \cdots, n.
	$
	We define the response vector as $\y = (y_1, y_2, \cdots, y_n)^T$ and the design matrix as $\X=(\x_1, \x_2, \cdots, \x_n)^T$. Under the classical setup when $n>p$, the OLS estimate of $\bbeta$ is obtained by minimizing the square error loss function $||\y - \X \bbeta||^2$, where $||\cdot||$ is the $L_2$ norm. The solution is $\hat{\bbeta} = (\X^T\X)^{-1} \X^T \y$, which is also the maximum likelihood estimator (MLE) of $\bbeta$. 
	
	Let $\mathcal{A} = \{j: 0\leq j \leq p, \beta_j \neq 0\}$ be the set of indices where $\bbeta$ has non-zero coefficients. In the true model, if there are $p_0$ non-zero  coefficients, then $p_0 = |\mathcal{A}|$, the cardinality of $\mathcal{A}$. Without loss of generality, we assume that $\beta_j\neq 0$ for $j\leq p_0$ and $\beta_j= 0$ for $j> p_0$. The OLS estimator is unbiased for $\bbeta$, but in small or moderate sample sizes when $p_0 < p$, it often has a large variance. On the other hand, shrinking or setting some regression coefficients to zero may improve the prediction accuracy. In this case, we may incorporate a small bias, but a greater reduction in the variance term is achieved. Thus, it often improves the overall mean square error (MSE). 
	
	We assume that the design matrix $\X$ is standardized so that $\sum_i x_{ij}/n=0$ and $\sum_i x^2_{ij}/n =1$ for all $j=2,3,\cdots, p$, where $x_{ij}$ is the $(i,j)$-th element of $\X$. 
	Parameter shrinkage is imposed by considering a penalized loss function
	\be 
	L(\bbeta| \lambda_n) = \frac{1}{2n} ||\y - \X \bbeta ||^2 + \sum_{j=1}^p P_{\lambda_n}(|\beta_j|),
	\label{loss_penelty}
	\ee    
	where the penalty function $P_{\lambda_n}(\cdot)$,  indexed by  regularized parameter $\lambda_n>0$, controls the model complexity. We assume that $P_{\lambda_n}(t)$ is non-decreasing function in $t$ and has a continuous derivative $P'_{\lambda_n}(t)=(\partial/\partial t) P_{\lambda_n}(t)$ in $(0,\infty)$. \cite{MR1157714} have shown that, under further assumption $P'_{\lambda_n}(0^+)>0$, the minimizer of Equation (\ref{loss_penelty}) has variable selection feature with zero components. 
	
	In general, 
	$P_{\lambda_n}(\cdot) = \lambda_n P(\cdot)$, where $\lambda_n$ balances between the bias and variance of the estimators. For example, in LASSO  $P_{\lambda_n}(|\beta|) = \lambda_n |\beta|$. More predictors are included as $\lambda_n \rightarrow 0^+$, producing smaller bias, but higher variance. For $\lambda_n=0$, we get the OLS estimate. On the other hand,
	fewer predictors stay in the model as $\lambda_n$ increases, and finally, only the intercept parameter remains  when $\lambda_n$ is larger than a threshold, say $\lambda_n > \lambda_0$. Therefore, with a properly tuned $\lambda_n$, the optimum prediction accuracy is achieved. 

	\section{Proposed Robust Penalized Regression} \label{sec:robust_reg}
	The density power divergence (DPD) measure between the model density $f_\btheta$ with parameter $\btheta \in \Theta$ and the empirical (or true) density $g$ is defined as
	\be 
	d_\alpha(f_\btheta, g) = 
	\left\{
	\begin{array}{ll}
		\int_y\left\{ f^{1+\alpha}_\btheta(y)-\left( 1+\frac{1}{\alpha}\right) f^{\alpha }_\btheta(y)g(y)+%
		\frac{1}{\alpha}g^{1+\alpha}(y)\right\} dy, & \text{for}\mathrm{~}\alpha>0, \\%
		[2ex]
		\int_y g(y)\log\left( \displaystyle\frac{g(y)}{f_\btheta(y)}\right) dy, & \text{for}%
		\mathrm{~}\alpha=0,%
	\end{array}
	\right. 
	\label{dpd}
	\ee
	where $\alpha$ is a tuning parameter \citep{MR1665873}. 	For $\alpha=0$, the DPD is obtained as a limiting case of $\alpha \rightarrow 0^+$; and the measure is called the  Kullback-Leibler divergence. 
	Given a parametric model, we estimate $\btheta$ by minimizing the DPD measure with respect to $\btheta$ over its parametric space $\Theta$. We call the estimator as the minimum power divergence estimator (MDPDE). 
	For $\alpha=0$, it is equivalent to maximize the log-likelihood function. Thus, the MLE is a special case of the MDPDE. The tuning parameter $\alpha$ controls the trade-off between  efficiency and robustness of the MDPDE -- robustness measure increases if $\alpha$ increases, but at the same time efficiency  decreases. 
	
	Let $\btheta = (\bbeta^T, \sigma^2)^T$ be the parameter of the regression model defined in Equation (\ref{reg_model}).
	The probability density function (pdf) of $y_i$, denoted by $f_\btheta(y_i|\x_i)$ or in short $f_i$, is given by
	\be
	f_i \equiv f_\btheta(y_i|\x_i) = \frac{1}{\sqrt{2\pi}\sigma} \exp^{-\frac{1}{2\sigma^2} (y_i - \x_i^T \bbeta )^2 }, \ \ \ i = 1, 2, \cdots, n.
	\label{fi}
	\ee
%
%
%
	Suppose data is centered and scaled in the pre-processing step. Although, the penalty function does not involve $\sigma$, but for notational simplicity, we denote the penalty function by $P_{\lambda_n}(\btheta) = \sum_{j=1}^p P_{\lambda_n}(|\beta_j|)$. It is obvious that the classical penalized regression analysis does not produce robust estimators due to the square error loss function in Equation (\ref{loss_penelty}). Therefore,  we propose a modified penalized loss function using the DPD measure as  
	\be 
	L_\alpha(\btheta|\X, \lambda_n) = \frac{1}{n}\sum_{i=1}^n d_\alpha(f_i, g_i) +  P_{\lambda_n}(\btheta), \label{cont}
	\ee 
	where $g_i \ i=1, 2, \cdots, n$ are the empirical probability density functions. As we are concern about the robustness properties of  estimators, data should be centered and scaled using robust statistics, such as the median and the mean absolute deviation (MAD).  
	For $\alpha>0$, the loss function in Equation (\ref{cont}) is simplified as
	\be 
	L_\alpha(\btheta|\X, \lambda_n) = \frac{1}{n}\sum_{i=1}^n V_i(\btheta|\X, \lambda_n, \alpha) + P_{\lambda_n}(\btheta) + c(\alpha) , \label{cont1}
	\ee 
	where $c(\alpha) = \frac{1}{\alpha} \int_y g^{1+\alpha}(y) dy$,  the third term of Equation (\ref{dpd}), is free of $\btheta$ and 
	\be
	V_i(\btheta|\X, \lambda,\alpha) = \frac{1}{(2\pi)^{\frac{\alpha}{2}} \sigma^\alpha \sqrt{1 + \alpha}} - \frac{1+\alpha}{ \alpha}  f_i^\alpha . \label{vi}
	\ee
	The MDPDEs of $\bbeta$ and $\sigma$ are obtained by minimizing $L_\alpha(\btheta|\X, \lambda_n)$ over $\bbeta \in \mathcal{R}^{p+1}$ and $\sigma >0$. If the $i$-th observation is an outlier, the value of $f_i$ is very small compared to other samples. When $\alpha>0$, the second term of Equation (\ref{vi}) is negligible for that $i$, thus the resulting MDPDE becomes robust against outlier. On the other hand, when $\alpha=0$, we have $V_i(\btheta|\X, \lambda_n,\alpha) = -\log(f_i)$; and it diverges as $f_i \rightarrow 0$. So, the MLE breaks down in the presence of outliers as they dominate the loss function. 
	
	\section{Computation Algorithm} \label{sec:algo}
	Let us define
	\be
	\nabla V(\bbeta) = \frac{1}{n}\sum_{i=1}^n \frac{\partial }{\partial \bbeta} V_i(\btheta|\X, \lambda_n,\alpha)  = - \frac{1+\alpha}{n}\sum_{i=1}^n \bu_i f_i^\alpha,
	\label{2nd_diff}
	\ee
	where the score function
	$ 
	\bu_i = \frac{\partial}{\partial \bbeta} \log f_i = 
	\frac{(y_i - \x_i^T \bbeta )}{\sigma^2} \x_i,
	$
	and $f_i$ is given in Equation (\ref{fi}). Note that $\nabla V(\bbeta)$  depends on $\X, \lambda_n,\alpha, \bbeta$ and $\sigma$, but for simplicity in the notation, we wrote it as a function of $\bbeta$ only. 
	The estimating equations for MDPDEs of $\bbeta$ and $\sigma$ are given by:
	\begin{align} 
	- \frac{1+\alpha}{n}\sum_{i=1}^n \bu_i f_i^\alpha
	+ P'_{\lambda_n}(\bbeta) &= 0,
	\label{est_beta}\\
	-\frac{\alpha}{(2\pi)^{\alpha/2} \sigma^\alpha \sqrt{1+\alpha}} + \frac{1+\alpha}{n}  \sum_{i=1}^n &\left\{1 - \frac{(y_i - \x_i^T \bbeta )^2}{\sigma^2} \right\} f_i^\alpha = 0,
	\label{est_sigma}
	\end{align}
	where $P'_{\lambda_n}(\bbeta) = \frac{\partial}{\partial \bbeta} P_{\lambda_n}(\btheta) = \sum_{j=1}^p P'_{\lambda_n}(|\beta_j|)$. Equations (\ref{est_beta}) and (\ref{est_sigma}) contain a system of $(p+2)$ non-linear equations, which may be difficult to solve. Following \cite{MR2395832}, we approximate the first term of the loss function in Equation (\ref{cont1}) by a quadratic function of $\bbeta$. Differentiating Equation (\ref{2nd_diff}) we get
	\be 
	\nabla^2 V(\bbeta) = - \frac{1+\alpha}{n}\sum_{i=1}^n \left(  \alpha \bu_i \bu_i^T f_i^\alpha +  \nabla \bu_i  f_i^\alpha \right),
	\label{2nd}
	\ee
	where $  \nabla \bu_i  = - \frac{1}{\sigma^2} \x_i \x_i^T $. Now, $\nabla^2 V(\bbeta)$ is a positive semi-definite matrix and can be decomposed as $\Z^T \Z$, where $\Z$ is a $(p+1)\times (p+1)$ matrix. Let us define $\boldsymbol{Y}^* = (\Z^T)^{-1} ( \nabla^2 V(\bbeta) \bbeta - \nabla V(\bbeta) )$. Then, the loss function in Equation (\ref{cont1}) can be approximated by:
	\be 
	L_\alpha(\btheta |\X, \lambda_n) \approx \frac{1}{2n} ||\boldsymbol{Y}^* - \Z\bbeta||^2 + P_{\lambda_n}(\btheta) + c(\alpha) . \label{cont2}
	\ee
	Therefore, the MDPDE of $\bbeta$ can be obtained iteratively  using the existing package of the corresponding penalized regression, eg., we may use the LARS algorithm in case of LASSO penalty  \citep{MR2060166}. The estimation procedure is given in Algorithm \ref{algo:MDPDE}.
	
	\begin{algorithm}
		\caption{Computation of MDPDEs of $\bbeta$ and $\sigma$}
		\begin{algorithmic}[1]
			\State Choose tuning parameters $\alpha$ and $\lambda_n$.
			\State {\bf Pre-processing:} Center and scale data using robust statistics.
			\State {\bf Initialization:} Initialize $\bbeta$ and $\sigma$ by OLS or any robust estimators.
			\While {convergence of the estimators of $\bbeta$ and $\sigma$}
			\State Compute $\nabla V(\bbeta), \nabla^2 V(\bbeta), Z$ and $\boldsymbol{Y}^*$.
			\State Update $\bbeta$ by minimizing Equation (\ref{cont2}).
			\State Update $\sigma$ by solving Equation (\ref{est_sigma}) or minimizing Equation (\ref{cont1}).
			\EndWhile
			\State {\bf Post-processing:} Unstandardize  $\bbeta$ and $\sigma$ by inverting pre-processing step.
		\end{algorithmic} \label{algo:MDPDE}
	\end{algorithm}

	\section{Asymptotic Distribution of the MDPDE} \label{sec:asymp}
	Suppose $g$ is the true data generating distribution, whereas $f_\btheta$ with $\btheta \in \Theta$ is the family containing the model distributions. We define $f_i = f_{\btheta}(\cdot|\x_i)$ and $g_i = g(\cdot|\x_i)$ for $i=1,2, \cdots, n$. Let $\btheta_g= (\bbeta_g^T, \sigma_g^2)^T$ be the  value of $\btheta$ that minimizes $\sum_i d_\alpha(f_i, g_i)$ over $\btheta \in \Theta$. In Section \ref{sec:robust_reg}, $g_i$ is referred to an empirical pdf, and the resulting minimizer of $\sum_i d_\alpha(f_i, g_i)$ produces the non-penalized MDPDE. Notice that $\btheta_g$ is the true value of the parameter if the model is correctly specified. However,  it is not necessary that $g$ is a member of the model family. In that case,  $f_{\btheta_g}$ is the closest density function to $g$ with respect to the DPD measure $\sum_i d_\alpha(f_i, g_i)$  over $\btheta \in \Theta$. 
	We assume that $\bbeta_g$ is sparse, and the set corresponding to the non-zero elements is given by $\mathcal{A} = \{j: 0\leq j \leq p, \beta_{gj} \neq 0\}$ where $|\A| = p_1 \leq p+1$.  Let us define $\bbeta_\mathcal{A}$ as the vector obtained from $\bbeta_g$ by selecting the elements corresponding to set $\mathcal{A}$. The remaining part of $\bbeta_g$ is called $\bbeta_{\bar{\A}}$. So, $\bbeta_{\bar{\A}}=\mathbf{0}$, the $(p+1-p_1)$-dimensional zero vector. 
	Suppose $\hat{\btheta} = (\hat{\bbeta}^T, \hat{\sigma}^2)^T$ is the MDPDE of $\btheta$ obtained by minimizing the loss function defined in Equation (\ref{cont1}). We also partition $\hat{\bbeta}$ as $\hat{\bbeta}_\A$ and  $\hat{\bbeta}_{\bar{\A}}$, where $\hat{\bbeta}_\A$ is a $p_1$-dimensional vector. Similarly, $\X$ is partitioned as $\X_\A$ and $\X_{\bar{\A}}$, where $\X_\A$ is a matrix of dimension $n\times p_1$. Let us define $\mathbf{\Sigma}_\A = \displaystyle { \lim_{n\rightarrow \infty} \frac{1}{n} \X_\A^T \X_\A}$ and
	\be 
	\xi_\alpha = (2\pi)^{-\frac{\alpha}{2}} \sigma^{-(\alpha+2)} (1+\alpha)^{-\frac{3}{2}} \mbox{ and } 
	\eta_\alpha = \frac{1}{4}(2\pi)^{-\frac{\alpha}{2}} \sigma^{-(\alpha+4)} \frac{2 + \alpha^2}{(1+\alpha)^{\frac{5}{2}}} .
	\label{xi}
	\ee
	
	The asymptotic distribution of the non-penalized MDPDE is derived in \cite{MR3117102}  without assuming a sparse representation of the true model. \cite{MR2796868} derived the asymptotic distribution of an M-estimator where the dimension ($p_n$) of  predictors increases over the sample size. The MDPDE is a special case of the M-estimator, but here we assume that the dimension of  predictors is fixed. To derive the asymptotic distribution of the MDPDE we require assumptions (A1)--(A7) of \cite{MR3117102} and  some selected assumptions from \cite{MR2796868} as follows: 
	
	\begin{itemize}
				
		\item[(C1)] $\max_j\{P'_{\lambda_n}(\beta_j): j \in \A\} = O(n^{-1/2})$, where $\bbeta_\A=(\beta_1, \beta_1, \cdots, \beta_{p_1})^T$. 
		
		\item[(C2)] $\max_j\{P''_{\lambda_n}(\beta_j): j \in \A\} \rightarrow 0$ as $n \rightarrow \infty$. 
		
		\item[(C3)]  $\displaystyle \liminf_{n\rightarrow \infty} \liminf_{\beta \rightarrow 0+} P'_{\lambda_n}(\beta)/\lambda_n > 0.$
		
		\item[(C4)] There exist two constants $C$ and $D$ such that $|P''_{\lambda_n}(\beta_1) - P''_{\lambda_n}(\beta_2)| \leq D|\beta_1 - \beta_2|$, if $\beta_1, \beta_2 > C\lambda_n$. 
		
		\item[(C5)] Let $d_n^2 = \max_i \x_i^T {\mathbf S}_n^{-1} \x_i$ where ${\mathbf S}_n = \X^T \X$. For large $n$, there exists a constant $s>0$ such that $d_n \leq s n^{-1/2}$. 
				
	\end{itemize}

	\begin{theorem}
		
		Assume that the regularity conditions  (A1)--(A7) of \cite{MR3117102} and (C1)--(C5) hold. Then, the  asymptotic distributions of the MDPDEs $\hat{\bbeta} = (\hat{\bbeta}_\A, \hat{\bbeta}_{\bar{\A}})^T$ and $\hat{\sigma}^2$ have the following properties.
		\begin{enumerate}
			\item Sparsity: $\hat{\bbeta}_{\bar{\A}} = \mathbf{0}$ with probability tending to 1.
			\item Asymptotic Normality of $\hat{\bbeta}_\A$: $\sqrt{n} (\hat{\bbeta}_\A - \bbeta_\A ) \overset{a}{\sim} N\left(\bo{b}, \frac{\xi_{2\alpha}}{\xi_{\alpha}^2} \mathbf{\Sigma}_\A^{-1}\right)$, where $\bo{b} = \frac{\sqrt{\xi_{2\alpha}}}{\xi_{\alpha}} \mathbf{\Sigma}_\A^{-1/2} \lim_{n\rightarrow \infty} P'_{\lambda_n}(\bbeta_\A)$.  
		\item Asymptotic Normality of $\hat{\sigma}^2$: $ 
		\sqrt{n} (\hat{\sigma}^2 - \sigma_g^2) \overset{a}{\sim} N(0, \sigma_\alpha^2),
		\mbox{  where } 
		\sigma_\alpha^2 = \frac{\eta_{2\alpha} - \frac{\alpha^2}{4} \xi_\alpha^2}{ \eta_\alpha^2} .
		$
		\item Independence: $\hat{\bbeta}_\A$ and $\hat{\sigma}^2$ are asymptotically independent. 
	\end{enumerate}
\label{theorem:asymp}
	\end{theorem}
The theorem ensures that,  for large sample sizes, our procedure correctly drops the variables that don't have any significant contribution to the true model. So, the method selects variables consistently. Moreover, the estimators of nonzero coefficients $(\hat{\bbeta}_\A)$ have the same asymptotic distribution as they would if the zero coefficients $(\bbeta_{\bar{\A}})$ were known in advance. But the penalized MDPDE is a biased estimator. This feature is also observed in other penalized estimators. An asymptotic unbiased estimator of $\bbeta_\A$ is $\hat{\bbeta}_\A - \bo{b}/\sqrt{n}$.

	One important use of the asymptotic distribution of the penalized MDPDE is in selecting the optimum value of the DPD parameter $\alpha$. 
	In practice,  $\alpha$ is chosen by the user depending on the desired level of robustness measure at the cost of efficiency. Alternatively, following  \cite{Warjones}, one may minimize  the  mean square error (MSE) of $\hat{\bbeta}_\A$ to obtain the optimum value of $\alpha$ adaptively. The  empirical estimate of the MSE, as the function of a pilot estimator $\bbeta^P_\A$, is given by
	\be
	\widehat{MSE}(\alpha) = (\hat{\bbeta}_\A - \bbeta^P_\A)^T (\hat{\bbeta}_\A - \bbeta^P_\A) +\frac{\xi_{2\alpha}}{\xi_{\alpha}^2}\tr(\X_\A^T \X_\A)^{-1}.
	\label{adaptive_alpha}
	\ee
	In particular, we recommend that a robust estimator, such as the Huber or Tukey's M-estimator, should be used as a pilot estimator. This method is implemented to calculate the optimum value of $\alpha$ for the real data example in Section \ref{sec:data}. 
	
	\section{Influence Function}\label{sec:inf}
	In this section, we present the influence function following the approach of Huber \citep{MR606374}. It measures the effect of extreme outliers on the estimator. Let $f_{\btheta},  \btheta \in \Theta$ be the family of the target densities, where $\btheta_g$ is the true value of $\btheta$. We denote $f_i = f_\btheta(\cdot|\x_i)$  for $i=1,2,\cdots, n$.  Suppose the true data generating distribution $g_\tau$ has $\tau$ proportion contamination from   $T$, where $T$ is either a fixed point or a random variable. Then, the true density given $\x_i$ is written as $g_{\tau, i} = (1-\tau)f_i + \tau \delta_{t_i}$, where $\delta_{t_i}$ is a point-mass density function of $T$ at $t_i$. Here, $t_i$ is a realization of $T$  for $i=1, 2, \cdots, n$.  Suppose the true value of the parameter $\btheta_g$ is now shifted to $\btheta_{\tau, g}$ due to contamination.  So, for large sample sizes,  $f_{\btheta_{\tau, g}}$ is the closest density function to $f_{\btheta_g}$ with respect to the DPD measure $\sum_i d_\alpha(f_i, g_{\tau, i})$  over $\btheta \in \Theta$.
	
	The influence function is defined by $IF(\btheta_g, \boldsymbol{t}) = \frac{\partial\btheta_{\tau, g}}{\partial \tau}|_{\tau = 0}$, where $\boldsymbol{t}=(t_1, t_2, \cdots, t_n)^T$. It gives the rate of asymptotic bias of an estimator to infinitesimal contamination in the distribution. A bounded influence function suggests that the corresponding estimator is robust against extreme outliers. Let us define
	\be
	\Psi_n = \left(
	\begin{array}{c c}
		\frac{\xi_\alpha}{n} \X^T \X & 0\\
		0 & \eta_\alpha
	\end{array}
	\right), \ \
	\Psi = \left(
	\begin{array}{c c}
		\xi_\alpha \bo{\Sigma} & 0\\
		0 & \eta_\alpha
	\end{array}
	\right),
	\label{omega}
	\ee
	where $\mathbf{\Sigma} = \displaystyle { \lim_{n\rightarrow \infty} \frac{1}{n} \X^T \X}$ and $\xi_{\alpha}$ and $\eta_\alpha$ are defined in Equation (\ref{xi}). The following theorem gives the influence function of the MDPDE.
	
	\begin{theorem}
	The influence function of the MDPDE for $\alpha>0$ is given by
	\be
	IF(\btheta_g, \boldsymbol{t}) = \left[\Psi_n + \frac{1}{1+\alpha}  P''_{\lambda_n}(\btheta_g) \right]^{-1}	\frac{1}{n}\sum_{i=1}^n  
	\left(
	\begin{array}{c}
		\frac{t_i - \x_i^T \bbeta_g }{(2\pi)^{\alpha/2}\sigma_g^{\alpha +2}} \exp\left[ - \frac{(t_i - \x_i^T \bbeta_g )^2}{2\sigma_g^2}\right]  \x_i\\
 \frac{(y_i - \x_i^T \bbeta_g )^2 - \sigma_g^2}{2(2\pi)^{\alpha/2}\sigma_g^{\alpha +4}}   \exp\left[ - \frac{(t_i - \x_i^T \bbeta_g )^2}{2\sigma_g^2}\right]	-\frac{\alpha }{2} \eta_\alpha
	\end{array}
	\right).
	\ee
		\label{theorem:influence}
\end{theorem} 
Under assumption (C2), $\displaystyle {\lim_{n\rightarrow \infty} \left[\Psi_n + \frac{1}{1+\alpha}  P''_{\lambda_n}(\btheta_g)\right] = \Psi}$. So, for large sample sizes, the penalty function does not have any role in the robustness of the MDPDE.  We observe that $IF(\btheta_g, \boldsymbol{t})$ is bounded for all $\alpha>0$ as $\exp(-x^2)$, $x\exp(-x^2)$ and $x^2\exp(-x^2)$ are bounded functions for $x \in \mathbb{R}$. For this reason, the penalized MDPDE of $\bbeta$ and $\sigma$ are robust against outliers. On the other hand, it is well known that the OLS estimator (corresponds to $\alpha=0$) is non-robust as its influence function is unbounded. In the simulation study, we further explore the robustness properties of the penalized MDPDE.

	\section{Robust Model Selection Criteria}\label{sec:selection}
	The model selection criterion plays a key role in choosing the best model for high-dimensional data analysis. 
	In a regression setting, it is well known that omitting an
	important explanatory variable may produce severe bias is  parameter estimates and prediction results. On the other hand, including unnecessary predictors may degrade the efficiency of the resulting estimation and yields less accurate prediction. Hence, selecting the best model based on a finite sample is always a problem of interest for both theory and application in this field. There are several important and widely used selection criteria, e.g. the Mallows's $C_p$ statistic \citep{mallows1973some},  the Akaike information criterion (AIC) \citep{MR0483125} etc. However, those selection criteria are based on the classical estimators, so they show very poor performance in the presence of heavy-tailed error and outliers. To overcome the deficiency, we propose robust versions of those methods to select the best sub-model by choosing the optimum value of regularization parameter $\lambda_n$. 
	
	\subsection{Robust $C_p$ Statistic and Degrees of Freedom} \label{sec_df}
	Suppose for the sub-model, the true selection set is given by $\mathcal{A} = \{j: 0\leq j \leq p, \beta_j \neq 0\}$. Let us define $\X_\mathcal{A}$ as the matrix obtained from $\X$ by selecting  columns corresponding to set $\mathcal{A}$. Similarly, $\hat{\bbeta}_\mathcal{A}$ and $\bbeta_\mathcal{A}$ are defined based on the set $\A$.  We further define
	\be
	J_\mathcal{A} = (\hat{\bbeta}_\mathcal{A} - \bbeta_\mathcal{A})^T \X_\mathcal{A}^T  \X_\mathcal{A} (\hat{\bbeta}_\mathcal{A} - \bbeta_\mathcal{A}).
	\ee
	Following \cite{mallows1973some}, we consider  $\frac{1}{\sigma^2}  E[J_\mathcal{A}]$ as a measure of prediction adequacy. Let $RSS_\mathcal{A}$ be the residual sum of squares for the sub-model. Then, if the sub-model is true, we have
	\begin{align}
	E(RSS_\mathcal{A}) &= E\left[(\y - \X_\mathcal{A} \hat{\bbeta}_\mathcal{A})^T (\y - \X_\mathcal{A} \hat{\bbeta}_\mathcal{A})\right]\\
	&= E\left[(\y - \X_\mathcal{A} \bbeta_\mathcal{A})^T (\y - \X_\mathcal{A} \bbeta_\mathcal{A})\right] - 2 E[(\y - \X_\mathcal{A} \bbeta_\mathcal{A})^T \X_\mathcal{A} (\hat{\bbeta}_\mathcal{A} - \bbeta_\mathcal{A})] \\
	& \ \ \ \ \ \ \ \ + E[(\hat{\bbeta}_\mathcal{A} - \bbeta_\mathcal{A})^T \X_\mathcal{A}^T  \X_\mathcal{A} (\hat{\bbeta}_\mathcal{A} - \bbeta_\mathcal{A})]\\
	& = n\sigma^2 - 2 \sigma^2 df + E(J_\mathcal{A}),
	\label{rcp}	
	\end{align}	
	where $df = \frac{1}{\sigma^2} E[(\y - \X_\mathcal{A} \bbeta_\mathcal{A})^T \X_\mathcal{A} (\hat{\bbeta}_\mathcal{A} - \bbeta_\mathcal{A})]$ is called the degrees of freedom or the ``effective number of
	parameters'' of the regression model. 
	
	\begin{lemma}
		If the sub-model is true, the degrees of freedom is expressed as
		\be  
		df = \frac{\xi_\alpha }{n}  \tr ( \X_\mathcal{A} \s^{-1} \X^T_\mathcal{A}) + o(1),
		\label{df}
		\ee
		where  
		\be  
		\s = \frac{\xi_\alpha }{n} \X^T_\mathcal{A} \X_\mathcal{A} + \frac{1}{1 + \alpha} P''_{\lambda_n}(\bbeta_\mathcal{A}),
		\label{A}
		\ee
		and $\xi_\alpha$ is given in Equation (\ref{xi}).
		\label{lemma_cp}
	\end{lemma}
The proof of the lemma is given in the supplementary materials.  Using this lemma, we estimate $\frac{1}{\sigma^2}  E[J_\mathcal{A}]$ from Equation (\ref{rcp}). We denote it by $RC_p$, the robust $C_p$ statistic. So, the $RC_p$ is given by
	\be 
	RC_p = \frac{1}{\hat{\sigma}_u^2} RSS_\mathcal{A} - n + \frac{2 \hat{\xi_\alpha} }{n}  \tr ( \X_\mathcal{A} \s^{-1} \X^T_\mathcal{A}),
	\ee
	where  $\hat{\xi_\alpha}$ is the estimate of $\xi_\alpha$ obtained from Equation (\ref{xi}). Here, $\hat{\sigma}_u$ is  a robust and unbiased estimator of $\sigma$ preferably using the full model where $\lambda_n=0$. The optimum value of the penalty parameter $\lambda_n$ is obtained by minimizing $RC_p$ using an iterative algorithm. Note that $RC_p$ is a function of $\lambda_n$ as it controls the selection set $\mathcal{A}$ and the estimates of the parameters. More specifically, $\hat{\xi_\alpha}$ and $RSS_\mathcal{A}$ are computed using the penalized MDPDE that involves $\lambda_n$. Now, $RSS_\mathcal{A}$ is outlier sensitive, so using Theorem \ref{theorem:asymp}, it is replaced by a consistent estimator $n \hat{\sigma}$, where  $\hat{\sigma}$ is the penalized MDPDE of $\sigma$ under the sub-model.   
	
	 Using assumption (C2), we have $\lim_{n\rightarrow \infty} \s = \xi_{\alpha} \bo{\Sigma}_\mathcal{A}$, where $\bo{\Sigma}_\mathcal{A}= \lim_{n\rightarrow \infty} \frac{1 }{n} \X^T_\mathcal{A} \X_\mathcal{A}$. So, asymptotically  the degrees of freedom  simplifies to $df=|\A|$, the number of non-zero regression coefficients. 
	Therefore, for large sample sizes
	$
	RC_p = \frac{n \hat{\sigma}}{\hat{\sigma}_u^2}  - n + 2|\A|,
	$
	which is equivalent to the classical $C_p$ statistic, but the estimators are replaced by suitable penalized MDPDE estimators.  
	
	\subsection{Robust AIC} \label{sec_raic}
	In this section, we assume that the true density $g$ belongs to the family of the model densities, i.e. $g = f_{\btheta_g}$ for some $\btheta_g \in \Theta$. Suppose the penalty function $P_{\lambda_n}(\cdot)$ in Equation (\ref{cont}) creates a sub-model where the set of non-zero elements of $\bbeta$ is given by $\mathcal{A} = \{j: 0\leq j \leq p, \beta_j \neq 0\}$ and $|\A| = p_1 \leq p+1$.  Let $\btheta_\A$ be the value of $\btheta \in \Theta_\A$ under the restricted sub-model that minimize $E(L_\alpha(\btheta|\X, \lambda_n))$. In this section, all expectations are calculated under the generating model $g$. Suppose $\hat{\btheta}_\A$ is the penalized MDPDE of $\btheta_\A$ for a given value of $\alpha$. For $\alpha=0$,  $\hat{\btheta}_\A$ is the MLE of $\btheta_\A$ and  $d_0(f_{\hat{\btheta}_\A}, f_{\btheta_g})$ becomes the Kullback-Leiber distance between two densities $f_{\hat{\btheta}_\A}$ and $f_{\btheta_g}$. The classical AIC minimizes the estimate of $E[d_0(f_{\hat{\btheta}_\A}, f_{\btheta_g})]$ assuming that $\btheta_\A$ lies very close to $\btheta_g$ (\citealp{MR1458291}). To make the procedure robust against outliers, we  minimize the estimate of $E[d_\alpha(f_{\hat{\btheta}_\A}, f_{\btheta_g})]$ using a suitable value of $\alpha$. For a random sample of size $n$, our goal is to find the optimum value of $\lambda_n$ that produces the best sub-model.
	
	Let us define
		\be
	\bsigma^*_\A = \left(
	\begin{array}{c c}
		\frac{\xi_{2\alpha}}{\xi_{\alpha}^2} \mathbf{\Sigma}_\A^{-1} & 0\\
		0 & \sigma_\alpha^2
	\end{array}
	\right), \ \
	\Psi_\A = \left(
	\begin{array}{c c}
		\xi_\alpha \bo{\Sigma}_\A & 0\\
		0 & \eta_\alpha
	\end{array}
	\right), \ \
	\bo{b}^* = \left(
	\begin{array}{c }
		\bo{b}\\
		0 
	\end{array}
	\right),
	\label{sigma_star}
	\ee
	 where $\bo{\Sigma}_\A, \ \sigma_\alpha, \ \xi_\alpha, \ \eta_\alpha$ and $\bo{b}$ are given in Theorem \ref{theorem:asymp}. Note that $\bsigma^*_\A$ and $\bo{b}^*$ are, respectively, the variance-covariance matrix and bias of $\sqrt{n}\hat{\btheta}_\A$. The following theorem gives the expression of the robust AIC. 
	
	\begin{theorem}
		Suppose the regularity conditions of Theorem \ref{theorem:asymp} hold true.  
	Then, the robust Akaike information criterion (AIC) is defined by 
	\be
	RAIC = 
	\left\{
	\begin{array}{ll}
			- \frac{1+\alpha}{ \alpha} \sum_{i=1}^n f_i^\alpha + \tr\left[(\hat{\bsigma}_\A^* + \hat{\bo{b}}^{*} \hat{\bo{b}}^{*T}) \left\{ \hat{\Psi}_\A +  P''_{\lambda_n}(\hat{\btheta}_\A) \right\}\right], & \text{for}\mathrm{~}\alpha>0, \\%
		[2ex]
			-  \sum_{i=1}^n \log(f_i) + \tr\left[(\hat{\bsigma}_\A^* + \hat{\bo{b}}^{*} \hat{\bo{b}}^{*T}) \left\{ \hat{\Psi}_\A +  P''_{\lambda_n}(\hat{\btheta}_\A) \right\}\right], & \text{for}%
		\mathrm{~}\alpha=0,%
	\end{array}
	\right. 
	\ee
where $\hat{\bsigma}_\A, \ \hat{\bo{b}}^*$, $\hat{\Psi}_\A$, $\hat{\xi}_{\alpha}$ and $\hat{\xi}_{2\alpha}$ are the estimates of $\bsigma_\A, \ \bo{b}^*$, $\Psi_\A$, $\xi_{\alpha}$ and $\xi_{2\alpha}$, respectively.
		\label{theorem:aic}
\end{theorem}
	 The derivation of the robust AIC is given in supplementary materials. For a given $\alpha$, we fit a set of candidate models by conducting a grid for $\lambda_n$ values in the log-scale over a suitable interval. The optimum value of the penalty parameter $\lambda_n$ is then  selected that minimizes the RAIC.

	\section{Simulation Study} \label{sec:simulatoin}
\begin{figure}
		\centering%
		\begin{tabular}{c}
			\includegraphics[height=7.5cm, width=15cm]{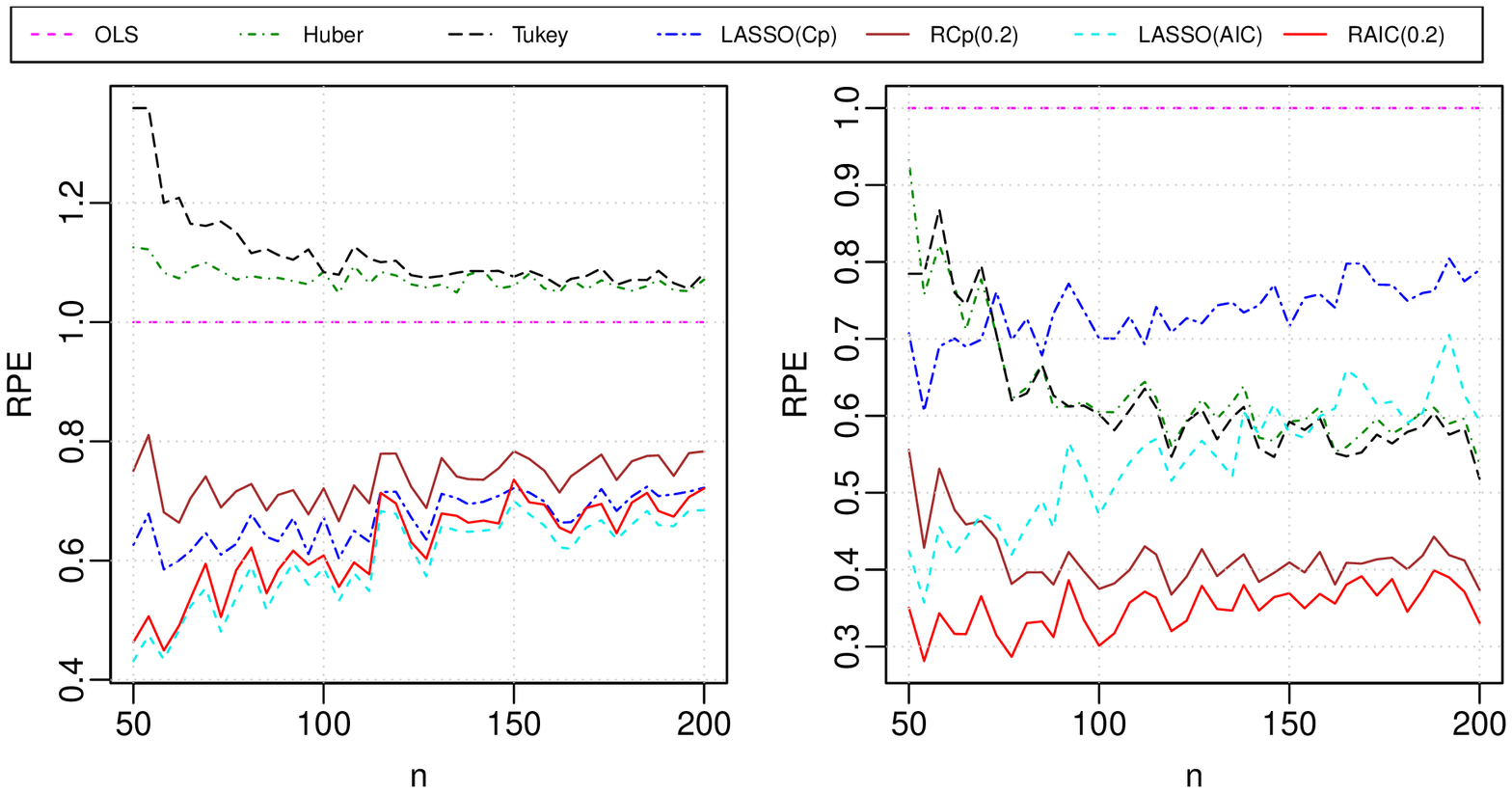}\\
			(a)  \hspace{2.5in}  (c)\\
			\includegraphics[height=7.5cm, width=15cm]{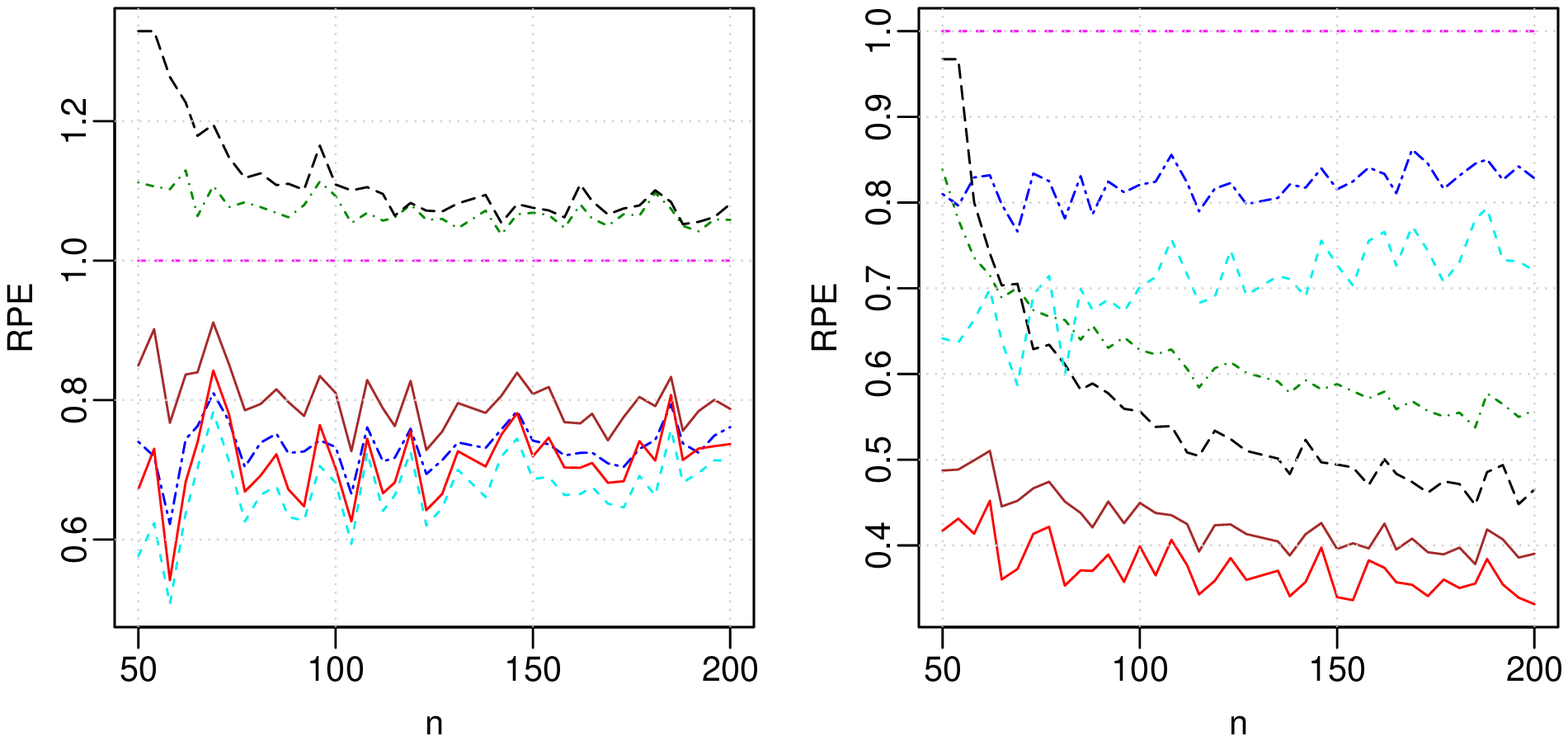} \\
			(b)  \hspace{2.5in}  (d)
		\end{tabular}
		\caption{The median of the estimated RPE relative to OLS for different estimators over 500 replications in the uncontaminated data (a and b), and data with (c)  1\% outliers at $\mu_c=10\sigma$ and (d) 5\% outliers at $\mu_c=5\sigma$. In (a) and (c) SNR=1 and in (b) and (d) SNR=10.}
		\label{fig:mse}
	\end{figure}

	We have presented an extensive simulation study to demonstrate the advantage of our proposed method. We simulated a data-set from the regression model given in Equation (\ref{reg_model}) with $p=25$ predictors. A sparse regression coefficient $\bbeta$ with 60\% null  components (i.e. $\beta_j=0$) is considered for this study. The value of the intercept parameter is fixed to $\beta_0=1$. At first, around half of the non-null regression coefficients are independently generated from the uniform distribution $U(1,2)$ and other half are taken from $U(-2,-1)$. Then, the regression coefficients are kept fixed for all simulations. The regressor variables are generated from a multivariate normal distribution, where each individual component $X_j$ follows the standard normal distribution. To introduce dependency among the regressor variables, the structure of the covariance matrix is taken as the first order auto-regressive model, AR(1), with correlation coefficient $\rho = 0.5$.   The values of $\sigma$, the standard deviation  of the error term in Model (\ref{reg_model}), are chosen such a way that the  signal-to-noise (SNR) ratios are 1 and 10 in Figure \ref{fig:mse} for plot (a) and (b), respectively.  We generated samples of sizes $n=50$ to $n=200$, and replicated the process 500 times for all $n$. In Equation (\ref{cont}), we considered the penalized DPD with $\alpha=0.2$ and the $L_1$ penalty function as used in LASSO. It should be mentioned here that, although we used only the $L_1$ penalty for the illustrative purpose, any other penalty function could be used in this example. Our method is very general and all the theoretical results are derived for an arbitrary penalty function provided the standard regularity conditions are satisfied. We also tested several values of $\alpha$, but for simplicity in the presentation, the results for $\alpha=0.2$ are reported here.  The optimum regularized parameter $\lambda$ is calculated based on the robust $C_p$ and AIC as discussed in Sections \ref{sec_df} and \ref{sec_raic}, respectively. These penalized MDPDEs are denoted by $RCp(0.2)$ and $RAIC(0.2)$, respectively. Our proposed method is compared with the classical LASSO estimators where the optimum $\lambda$ is selected using the classical  $C_p$ and AIC. For comparison, other than the OLS, we have taken two robust  (non-penalized) regression methods using the Huber and Tukey's M-estimators \citep{MR606374}. The default parameters in `rlm' package of R are used for these two estimators.  Once we calculated seven set of estimators based on a training data, we simulated another set of 1,000 test data to compare their performance using  $E[(\hat{y} - x^T\bbeta)^2]$, the relative prediction error (RPE). For each estimator and for each $n$, the  medians of the empirical RPE compared to the OLS are plotted in Figure \ref{fig:mse} (a) and (b), where $SNR=1$ and 10, respectively.  The both figures show that penalization technique increases the performance of these estimators in case of sparse regression. All the penalized estimators perform equally good in these situations, however, the methods using the AIC or RAIC are slightly better. This simulation study also demonstrates the well known fact that the prediction error of the classical robust estimators is higher than the OLS in pure data.

	In the next  simulation setup, we have contaminated $\tau\%$ outliers in the data, where the error term is generated from $\epsilon \sim N(\mu_c, 0.01)$ with $\mu_c$ being moved to a large value. Remaining $(100-\tau)\%$ data are generated using the first setup where $\epsilon \sim N(0, \sigma^2)$. So,  Model (\ref{reg_model}) is generated using a heavy-tailed error distribution. Other than outliers, all parameters were kept unchanged in this simulation. In the first contaminated case where $SNR=1$, we have taken $\mu_c=10\sigma$ and 1\% outliers, i.e. $\tau=0.01$. The plot for the estimates of the relative RPE is given in Figure \ref{fig:mse}(c). We observe that the performance of our robust estimators is far better than  other methods. In fact, the raw RPEs of all robust methods considered here are almost unchanged in the presence of outliers, whereas the performance of the OLS and LASSO estimators deteriorated significantly, and thus it creates a big difference in the relative performance. The plot also shows that our estimators have much smaller prediction errors compared to the Huber and Tukey's estimators. Moreover, these classical robust methods produce non-sparse coefficients, but the penalized MDPDEs have high values of specificity (around 80\%--100\% for RCp and 40\%--80\% for RAIC) and sensitivity (around 40\% for RCp and 60\% for RAIC). Similar result is obtained in the second contaminated case with  $SNR=10$ when there are 5\% outliers at $\mu_c=5\sigma$, see Figure \ref{fig:mse}(d). 
	
	Overall the simulation study shows that our penalized DPD based estimators outperform the classical robust methods and the LASSO in the presence of outliers in data. And in pure data without any outliers, their performance is quite competitive with the LASSO. It should be noted here that, for $L_1$ penalty, Algorithm \ref{algo:MDPDE} uses LASSO iteratively; thus it  can also handle high-dimensional data with $n\ll p$.

	\section{Real Data Analysis} \label{sec:data}     
	
	The data-set was obtained from a pilot grant funded by Blue Cross Blue Shield of Michigan Foundation.  The primary goal was to develop a pilot computer system (Intelligent Sepsis Alert) aimed towards increased automated sepsis detection capacity. We analyzed a subset of the data that is available to us. It contains 51 variables from 8,975 cases admitted to Detroit Medical Center (DMC) from 2014--2015.
	Both demographic and clinical data are available during the first six hours of patients' emergency department stay. The outcome variable ($y$) 	is the length of hospital stay. 
	Few variables are deleted as they had more than 75\% missing values. Some variables are also excluded as they contain texts mostly the  notes from the doctor or nurse.  We randomly partitioned the data-set into four equal parts where one sub-sample is used as the test set and remaining three sub-samples form the training set. We have calculated seven different estimators as used in the simulation study in Section \ref{sec:simulatoin}. For the penalized MDPDE, we considered several values of $\alpha$, however only the result using $\alpha=0.4$ is presented as the optimum $\alpha$ is around 0.4 in the full data. More precisely, we calculated the optimum value of $\alpha$ by minimizing the MSE in Equation (\ref{adaptive_alpha}). When the optimum penalty parameter $\lambda_n$ is obtained from the robust $C_p$ statistic, the optimum value of $\alpha$ becomes $0.3750$, and using the robust AIC it is $\alpha =0.3875$. 
	 The performance of the estimators is compared by the mean absolute prediction error (MAPE)  in the test data, where $\mbox{MAPE} = {\frac  {1}{n_t}}\sum _{{i=1}}^{n_t}|(y_i -\x_i^T\hat{\bbeta})/y_i|$,
	$n_t$ being the number observations of the test set. The process was replicated 100 times using different random partition of the data-set, then the mean MAPE relative to the OLS and the median percentage of dimension reduction are reported in Table \ref{tab:sepsis}. The result shows that our proposed methods reduce the MAPE by around 38\% compared to the OLS and classical LASSO. At the same time, unlike the Huber or Tukey's estimators, the penalized MDPDEs considerably reduce the dimension of predictor variables. 
	
	\begin{table}
		\centering
		\tabcolsep=0.1cm
		\small
		\begin{tabular}{l|ccccccc}
			\hline
			Estimators & OLS & Huber & Tukey & LASSO(Cp) & RCp(0.4) & LASSO(AIC) & RAIC(0.4) \\ 
			\hline
			Relative MAPE & 1.00 & 0.75 & 0.66 & 1.02 & 0.62 & 1.02 & 0.62 \\ 
			Dim. Reduction  & 0 & 0 & 0 & 22.86 & 28.57 & 25.71 & 34.29 \\ 
			\hline
		\end{tabular}
		\label{tab:sepsis}
		\caption{The mean of MAPE relative to the OLS and the median percentage of dimension reduction over 100 random resamples for different estimators for the sepsis data.}
	\end{table}

	The increased efficiency of the robust methods clearly indicates that the data-set contains a significant amount of outliers and they are not easy to detect for the high-dimensional data. For this reason, the multiple $R^2$ from the OLS method is just 11.42\%, but the value dramatically increased to 82.50\% and 85.32\% in case of Huber and Tukey's M-estimators, respectively. On the other hand, LASSO reduces 22.86\% and 25.71\% of variables, respectively, when the $C_p$ and the AIC are used. So, it reveals that several regressor variables do not contain any significant information for predicting the outcome variable $y$. Our proposed method successfully combines these two important properties -- the robustness property and a sparse representation of model. In our future research,  we would like to extend the robust penalized regression to the generalized linear model (GLM) so that it will be helpful to model the binary outcome of sepsis. 
	These methods could also be extended to obtain a suitable imputation method to deal with missing values.

	\section{Conclusion} \label{sec:conc}
		We have developed a robust penalized regression  method that can perform regression shrinkage and selection like the LASSO or SCAD, while being resistant to outliers or heavy-tailed errors like the LAD or quantile regression. The basic idea is to use an M-estimator based on the DPD measure  to estimate the model parameters, and then select the best model by using a suitable information criterion modified by the same robust estimators. A fast algorithm for the regression estimators is proposed that can be successfully  applied in the high-dimensional data analysis.  
		The asymptotic distribution and the influence function of the estimator are derived. Two robust information criteria are introduced by modifying the Mallow's $C_p$ and AIC to make the variable selection procedure stable against outliers. All the theoretical results are based on a generalized penalty function. So, using this procedure, one may robustify the classical penalized regression methods, such as LASSO, adaptive LASSO, SCAD, MCP, elastic net etc. The simulation studies as well as the real data example show improved performance of the proposed method over the classical procedures. Thus, the new procedure is expected to improve prediction power significantly for the high-dimensional  data where presence of outliers is very common.
		
	\bibliography{NSF_proposal_2018}

\clearpage

\appendix
\setcounter{page}{1}
\setcounter{section}{1}
\setcounter{equation}{0} 

\makeatletter\@addtoreset{equation}{section}
\def\theequation{\thesection.\arabic{equation}}

\begin{center}
	{\Huge{\bf Supplementary Materials}}
\end{center}

\subsection*{Proof of Theorem \ref{theorem:asymp}}	
The sparsity of the regression coefficient is directly proved from the Lemma 1 of \cite{MR2796868}. So,  $\hat{\bbeta}_{\bar{\A}}=\mathbf{0}$ with probability tending to 1. At the same time, for sufficiently large sample sizes, $\hat{\bbeta}_\A$ stays away from zero. Now, we  derive the asymptotic distribution of $(\hat{\bbeta}_\A, \hat{\sigma}^2)^T$ from the estimating equations (\ref{est_beta}) and (\ref{est_sigma}). Suppose the corresponding equations are written together as $ M_n(\hat{\btheta}) = 0$, where the first $p_1$ equations are obtained from (\ref{est_beta}) by taking equations corresponding to set $\A$, and  the last equation is  Equation (\ref{est_sigma}). Let $M_n^j(\hat{\btheta})$ be the $j$-th element of $M_n(\hat{\btheta}), \ j = 1, 2, \cdots,p_1+1$. We define $\A^* = \{\A, p+2\}$. Using a Taylor series expansion of $M_n^j(\btheta)$, we write
\be
M_n^j(\hat{\btheta}) = M_n^j(\btheta_g) + \sum_{k \in \A^*} (\hat{\theta}_k - \theta_{g,k}) M_n^{jk}(\btheta_g) + \frac{1}{2} \sum_{k \in \A^*} \sum_{l \in \A^*} (\hat{\theta}_k - \theta_{g,k}) (\hat{\theta}_l - \theta_{g,l}) M_n^{jkl}(\tilde{\btheta}_g),
\label{taylor}
\ee
where $\tilde{\btheta}_g$ belongs to the line segment connecting $\hat{\btheta}$ and $\btheta_g$. Here, $M_n^{jk}$ and $M_n^{jkl}$ are, respectively, the first and second order partial derivatives of $M_n^{j}$ with respect to the indicated components of $\btheta$. As $M_n^j(\hat{\btheta})=0$, we get
\be
\sum_{k \in \A^*} (\hat{\theta}_k - \theta_{g,k}) \left[ M_n^{jk}(\btheta_g) + \frac{1}{2} \sum_{l \in \A^*}  (\hat{\theta}_l - \theta_{g,l}) M_n^{jkl}(\tilde{\btheta}_g) \right]= - M_n^j(\btheta_g).
\label{taylor1}
\ee
Let us define $\hat{\btheta}^*=(\hat{\bbeta}_\A^T, \hat{\sigma}^2)^T$ and $\btheta_g^*=(\bbeta_\A^T, \sigma_g^2)^T$.  Combining terms for $j =1, 2, \cdots, p_1+1$, the above equation is written as
\be
\bo{A}_n (\hat{\btheta}^* - \btheta_g^*) =- M_n(\btheta_g),
\label{taylor11}
\ee
where the $(j, k)$-th element of $\bo{A}_n$ is
\be
A_{n, j, k} = M_n^{jk}(\btheta_g) + \frac{1}{2} \sum_{l \in \A^*}  (\hat{\theta}_l - \theta_{g,l}) M_n^{jkl}(\tilde{\btheta}_g).
\ee
We define
\be
\Psi_\A = \left(
\begin{array}{c c}
	\xi_\alpha \bsigma_\A & 0\\
	0 & \eta_\alpha
\end{array}
\right), \ \
\Omega_\A = \left(
\begin{array}{c c}
	\xi_{2\alpha}\bsigma_\A & 0\\
	0 & \eta_{2\alpha} - \frac{\alpha^2}{4} \xi_\alpha^2
\end{array}
\right),
\label{omega_new}
\ee
where  $\xi_{\alpha}$ and $\eta_\alpha$ are defined in Equation (\ref{xi}).
A direct calculation shows that $E(M_n(\btheta_g))=\bo{b}_1$ and $n V(M_n(\btheta_g)) = (1+\alpha)^2 \Omega_\A + o_p(\mathbf{1})$, where $\bo{b}_1=(\bo{b}_2^T, 0)^T$ and $\bo{b}_2 = \lim_{n\rightarrow \infty} P'_{\lambda_n}(\bbeta_\A)$. Moreover, using the central limit theorem (CLT), we get 
\be
\frac{\sqrt{n}}{1+\alpha} \Omega_\A^{-\frac{1}{2}} M_n(\btheta_g) \overset{a}{\sim} N(\bo{b}_1, I_{p+2}).
\ee
Therefore, from (\ref{taylor11}), we have
\be
\frac{\sqrt{n}}{1+\alpha} \Omega_\A^{-\frac{1}{2}} \bo{A}_n (\hat{\btheta}^* - \btheta_g^*) \overset{a}{\sim} N(\bo{b}_1, I_{p+2}).
\label{conv}
\ee
Using the week law of large numbers (WLLN), it can be shown that, under condition (C2)
\be
\frac{1}{1+\alpha}  \bo{A}_n  - \Psi_\A \overset{P}{\rightarrow} 0,
\label{an_conv}
\ee
where $\Psi_\A$ is defined in Equation (\ref{omega_new}). So, from Equation (\ref{conv}), we find
\be
\sqrt{n} \Omega_\A^{-\frac{1}{2}}  \Psi_\A  (\hat{\btheta}^* - \btheta_g^*) \overset{a}{\sim} N(\bo{b}_1, I_{p+2}).
\ee
Thus, the theorem is proved by collecting the corresponding components of $\hat{\btheta}^*$.

\subsection*{Proof of Theorem \ref{theorem:influence}}
The loss function in Equation (\ref{cont}) is written as
\begin{align}
L_\alpha(\btheta|\X, \lambda_n) &= \frac{1}{n}\sum_{i=1}^n d_\alpha(f_i, g_{\tau, i}) +  P_{\lambda_n}(\btheta)\\
&= \frac{1}{n}\sum_{i=1}^n \int \left[f_i^{1+\alpha} - \left(1 + \frac{1}{\alpha}\right) f_i^\alpha g_{\tau, i}\right]dy_i +  P_{\lambda_n}(\btheta) + c(\alpha),
\label{cont_inf}
\end{align}
where $c(\alpha)$ is given in Equation (\ref{cont1}). $L_\alpha(\btheta|\X, \lambda_n)$ is minimized at $\btheta=\btheta_{\tau, g}$. So, for $\alpha>0$, the estimating equation at $\btheta=\btheta_{\tau, g}$ becomes
\be
\frac{1}{n}\sum_{i=1}^n \int \left[f_i^{1+\alpha} \bu_i^* -  f_i^\alpha \bu_i^* g_{\tau, i}\right]dy_i + \frac{1}{1+\alpha}  P'_{\lambda_n}(\btheta) = 0, \label{inf_est}
\ee
where
\be 
\bu_i^* = \frac{\partial}{\partial \btheta} \log f_i = 
\left(
\begin{array}{c}
	\frac{y_i - \x_i^T \bbeta }{\sigma^2} \x_i\\
	\frac{(y_i - \x_i^T \bbeta )^2}{2\sigma^4} -\frac{1}{2\sigma^2}
\end{array}
\right).
\label{ui_star}
\ee
Note that both $f_i$ and $\bu_i^*$ are functions of $\btheta$. Differentiating  Equation (\ref{inf_est}) with respect to $\tau$, we get
\begin{align}
\frac{1}{n}\sum_{i=1}^n \int &\Big[\left\{(1+\alpha) f_i^{1+\alpha} \bu_i^* \bu_i^{*T} + f_i^{1+\alpha} \nabla \bu_i^* - \alpha f_i^\alpha \bu_i^* \bu_i^{*T} g_{\tau, i} -  f_i^\alpha \nabla \bu_i^* g_{\tau, i}\right\}\frac{\partial\btheta}{\partial \tau}  \\
& +  f_i^\alpha \bu_i^*  g_i - f_i^\alpha \bu_i^*  \delta_{t_i}
\Big]dy_i \Bigg|_{\btheta=\btheta_{\tau, g}} + \frac{1}{1+\alpha}  P''_{\lambda_n}(\btheta) \frac{\partial\btheta}{\partial \tau}\Bigg|_{\btheta=\btheta_{\tau, g}} = 0,
\label{inf}
\end{align}
where  $\nabla\bu_i^* = \frac{\partial^2}{\partial \btheta^T \partial \btheta} \log f_i$. Now, it can be shown that 
\be
\Psi_n = \lim_{\tau \rightarrow 0}\frac{1}{n}\sum_{i=1}^n \int \left[(1+\alpha) f_i^{1+\alpha} \bu_i^* \bu_i^{*T} + f_i^{1+\alpha} \nabla \bu_i^* - \alpha f_i^\alpha \bu_i^* \bu_i^{*T} g_i -  f_i^\alpha \nabla \bu_i^* g_i\right]dy_i,
\ee
where $\Psi_n$ is given in Equation (\ref{omega}). 
Therefore, rearranging the terms of Equation (\ref{inf}) and taking limit as $\tau \rightarrow 0$, we get
\be
IF(\btheta_g, \boldsymbol{t}) = \left[\Psi_n + \frac{1}{1+\alpha}  P''_{\lambda_n}(\btheta_g) \right]^{-1}	\frac{1}{n}\sum_{i=1}^n  \left[f_i^\alpha \bu_i^*|_{y_i =  t_i} - \int f_i^\alpha \bu_i^*  g_i dy_i \right],
\ee
where $f_i$ and $\bu_i^*$ are evaluated at $\btheta= \btheta_g$. A direct calculation shows that
\be
\int f_i^\alpha \bu_i^*  g_i dy_i = \left(
\begin{array}{c}
	\boldsymbol{0}\\
	-\frac{\alpha }{2} \eta_\alpha
\end{array}
\right),
\ee
where $\boldsymbol{0}$ is a null vector of length $p+1$. Thus, the final form of $IF(\btheta_g, \boldsymbol{t})$ is obtained using the expressions of $f_i$ and $\bu_i^*$ from Equation (\ref{fi}) and (\ref{ui_star}), respectively.

\subsection*{Proof of Lemma \ref{lemma_cp}}
In the proof of Theorem \ref{theorem:asymp}, we used a Taylor series expansion of the estimating Equations (\ref{est_beta}) and (\ref{est_sigma}) with respect to $\bbeta_\A$ and $\sigma^2$. Here, we expand only Equation (\ref{est_beta}) with respect to $\bbeta_\A$ treating $\sigma$ as constant. For notational simplicity, we define $\nabla V(\bbeta_\A)$ as the vector obtained by taking elements from Equation (\ref{2nd_diff}) corresponding to set $\A$. Similarly, $\nabla^2 V(\bbeta_\A)$ is defined from Equation (\ref{2nd}). Then, similar to Equation (\ref{taylor11}), we write
\be
\bo{A}_n^* (\hat{\bbeta}_\A - \bbeta_\A) = M_n^*(\bbeta_\A),
\label{taylor111}
\ee
where 
\begin{align}
 	M_n^*(\bbeta_\A) &= -\nabla V(\bbeta_\A) - P'_{\lambda_n}(\bbeta_\A),\\
 	\bo{A}_n^* &= E\left[\nabla^2 V(\bbeta_\A)\right] + P''_{\lambda_n}(\bbeta_\A) + o_p(1).
 	\label{an}
\end{align}
Suppose $\x_{\A i}$ is the obtained from $\x_i$ by selecting elements corresponding to $\A$.  From Model (\ref{reg_model}), we have $ \epsilon_i = \y_i - \x_i^T \bbeta_g = \y_i - \x_{\A i}^T \bbeta_\A$. Let us define 	$ 
 \bu_{\A i} = \frac{\partial}{\partial \bbeta_\A} \log f_i = 
 \frac{(y_i - \x_{\A i}^T \bbeta_\A )}{\sigma^2} \x_{\A i},
 $ for $i=1,2, \cdots, n$. So, we write
\begin{align}
E[\nabla^2 V(\bbeta_\A)] &= - \frac{1+\alpha}{n}\sum_{i=1}^n E\left[\left(  \alpha \bu_{\A i} \bu_{\A i}^T  +  \nabla \bu_{\A i}   \right)f_i^\alpha\right]\\
&= - \frac{1+\alpha}{n}\sum_{i=1}^n E\left[\left(  \frac{ \alpha\epsilon_i^2}{\sigma^4}  - \frac{1}{\sigma^2} \right) \x_{\A i} \x_{\A i}^Tf_i^\alpha\right]\\
&= - \frac{1+\alpha}{n} \left(\frac{1}{\sqrt{2\pi}\sigma}\right)^\alpha \X_\A^T \X_\A E\left[\left(  \frac{ \alpha\epsilon^2}{\sigma^4}  - \frac{1}{\sigma^2} \right)  \exp\left(-\frac{\alpha\epsilon^2}{2\sigma^2}\right)\right]\\
&= - \frac{1+\alpha}{n} \left(\frac{1}{\sqrt{2\pi}\sigma}\right)^\alpha \X_\A^T \X_\A \int_{\mathbb{R}}\left[\left(  \frac{ \alpha\epsilon^2}{\sigma^4}  - \frac{1}{\sigma^2} \right) \frac{1}{\sqrt{2\pi}\sigma} \exp\left(-\frac{(1+\alpha)\epsilon^2}{2\sigma^2}\right)\right]d\epsilon \\
&= - \frac{\sqrt{1+\alpha}}{n} \left(\frac{1}{\sqrt{2\pi}\sigma}\right)^\alpha \X_\A^T \X_\A \left(  \frac{ \alpha}{\sigma^2 (1+\alpha)}  - \frac{1}{\sigma^2} \right)  \\
&= \frac{(1+\alpha)\xi_\alpha}{n}\X_\A^T \X_\A,
\label{2nd1}
\end{align}
where $\xi_\alpha$ is defined in Equation (\ref{xi}). 
So, from Equation (\ref{an}), we write $\bo{A}_n^* = (1 + \alpha) \s + o_p(1).$ Now, from Equation (\ref{taylor111}), we get
\be
(\hat{\bbeta} - \bbeta_\A) =- \frac{1}{1 + \alpha} \s^{-1} M_n^*(\bbeta_\A) + o_p\left(1\right).
\label{beta}
\ee
Let us define $\boldsymbol{\epsilon} = (\epsilon_1, \epsilon_2, \cdots, \epsilon_n)^T$ and $\boldsymbol{B}_\A =\X_\A \s^{-1}$. Suppose $b_{ij}$ is the $(i,j)$-th element of $\boldsymbol{B}_\A$ and $m_i$ is the $i$-th element of $M_n^*(\bbeta_\A)$ for $i, j=1,2,\cdots, n$. Then, from Equation (\ref{beta}), we get
\begin{align}
E[(\y & - \X_\A \bbeta_\A)^T \X_\A (\hat{\bbeta}_\A - \bbeta_\A)] =- \frac{1}{1 + \alpha} E[\boldsymbol{\epsilon}^T \X_\A   \s^{-1} M_n^*(\bbeta_\A)] + o(1)\\
&=- \frac{1}{1 + \alpha} E\left[\sum_{i,j} \epsilon_i b_{ij} m_j \right]+ o(1)\\
&= \frac{1}{n\sigma^2}E\left[\sum_{i,j} \epsilon_i b_{ij} \left(\sum_{k} \epsilon_k x_{kj} f_k^\alpha + P'_{\lambda_n}(\bbeta_\A)\right) \right]+ o(1)\\
&= \frac{1+\alpha}{n\sigma^2} E\left[\sum_{i,j} \epsilon_i^2 b_{i,j}  x_{ij} f_i^\alpha \right]+ o(1)\\
&= \frac{1}{n\sigma^2} E(\epsilon^2f^\alpha)\sum_{i,j}  b_{ij}  x_{ij}  + o(1)\\
&= \frac{1}{n\sigma^2}  \tr(\boldsymbol{B}_\A \X_\A^T)  \left(\frac{1}{\sqrt{2\pi}\sigma}\right)^\alpha \int_{\mathbb{R}} \frac{\epsilon^2}{\sqrt{2\pi}\sigma} \exp\left(-\frac{(1+\alpha)\epsilon^2}{2\sigma^2}\right) d\epsilon  + o(1)\\
&= \frac{1}{n\sigma^2}  \tr(\X \s^{-1} \X_\A^T)  \left(\frac{1}{\sqrt{2\pi}\sigma}\right)^\alpha \frac{1}{\sqrt{1+\alpha}} \frac{\sigma^2}{1+\alpha} + o(1)\\
&= \frac{\xi_\alpha \sigma^2}{n}  \tr(\X_\A \s^{-1} \X_\A^T)   + o(1).
\end{align}


\subsection*{Proof of Theorem \ref{theorem:aic}}
Let us define $D_\alpha(\btheta|\X, \lambda_n) = \frac{1}{n}\sum_{i=1}^n d_\alpha(f_i, g_i)$. Note that 
\be
E[d_\alpha(f_{\hat{\btheta}_\A}, f_{\btheta_g})] = E[E[D_\alpha(\btheta |\X, \lambda_n) | \btheta = \hat{\btheta}_\A]]
\ee
Now, from Equation (\ref{cont}), we get
  	\be 
L_\alpha(\btheta|\X, \lambda_n) =D_\alpha(\btheta |\X, \lambda_n) +  P_{\lambda_n}(\btheta). \label{cont_new}
\ee 
As $L_\alpha(\btheta|\X, \lambda_n)$ is minimized at $\btheta = \hat{\btheta}_\A$ for $\btheta \in \Theta_R$, we have
\be
\left[\frac{\partial}{\partial \btheta} L_\alpha(\btheta |\X, \lambda_n) \right]_{\btheta = \hat{\btheta}_\A} =0. \label{est_der}
\ee
Moreover,  $E[L_\alpha(\btheta|\X, \lambda_n)]$ is minimized at $\btheta = \btheta_\A$ for $\btheta \in \Theta_R$, so
\be
E\left[\left[\frac{\partial}{\partial \btheta} L_\alpha(\btheta |\X, \lambda_n)\right]_{\btheta = \btheta_\A}\right] =0. \label{true_der}
\ee 
Using Theorem \ref{theorem:asymp}, we obtain from Equation (\ref{an_conv})
\be
\left[\frac{\partial^2}{\partial \btheta^T \partial \btheta} L_\alpha(\btheta |\X, \lambda_n) \right]_{\btheta = \hat{\btheta}_\A} = \left[\frac{\partial^2}{\partial \btheta^T \partial \btheta} L_\alpha(\btheta |\X, \lambda_n) \right]_{\btheta = \btheta_\A} + o_p(1) = 
\Psi_\A + \frac{1}{1 + \alpha} P''_{\lambda_n}(\btheta_\A) +  o_p(1).
\label{2nd_der}
\ee 
Now, $nE[E[D_\alpha(\btheta |\X, \lambda_n) | \btheta = \hat{\btheta}_\A]]$ is written as follows
\begin{align}
  	nE[E[D_\alpha(\btheta |\X, \lambda_n) | \btheta = \hat{\btheta}_\A]] &= nE[D_\alpha(\hat{\btheta}_\A |\X, \lambda_n) ]    + n\left\{E[D_\alpha(\btheta_\A |\X, \lambda_n)] - E[D_\alpha(\hat{\btheta}_\A |\X, \lambda_n) ] \right\}\\
  		& \hspace{.4in}  + n\left\{E[E[D_\alpha(\btheta |\X, \lambda_n) | \btheta = \hat{\btheta}_\A]] - E[D_\alpha(\btheta_\A |\X, \lambda_n) ] \right\}\\
  		& = nE[D_\alpha(\hat{\btheta}_\A |\X, \lambda_n) ]  + n\left\{E[L_\alpha(\btheta_\A |\X, \lambda_n)] - E[L_\alpha(\hat{\btheta}_\A |\X, \lambda_n) ] \right\}\\
  		& \hspace{.4in}  + n\left\{E[E[L_\alpha(\btheta |\X, \lambda_n) | \btheta = \hat{\btheta}_\A]] - E[L_\alpha(\btheta_\A |\X, \lambda_n) ] \right\}. \label{expr}
\end{align}
A Taylor series expansion of $n E[L_\alpha(\btheta_g |\X, \lambda_n)]$ about $\btheta = \hat{\btheta}_\A$ and using Equations (\ref{est_der}) and (\ref{2nd_der}), we get
\begin{align}
n & E[L_\alpha(\btheta_\A |\X, \lambda_n)]  - 	nE[L_\alpha(\hat{\btheta}_\A |\X, \lambda_n) ] \\
&= nE\left[(\btheta_\A - \hat{\btheta}_\A) \left[\frac{\partial}{\partial \btheta} L_\alpha(\btheta |\X, \lambda_n) \right]_{\btheta = \hat{\btheta}_\A} \right]\\
& \hspace{1in} + \frac{n}{2}E\left[(\btheta_\A - \hat{\btheta}_\A)^T \left[\frac{\partial^2}{\partial \btheta^T \partial \btheta} L_\alpha(\btheta |\X, \lambda_n) \right]_{\btheta = \hat{\btheta}_\A} (\btheta_\A - \hat{\btheta}_\A)\right] + o(1)\\
&=  \frac{n}{2}E\left[(\btheta_\A - \hat{\btheta}_\A)^T   \left\{ \Psi_\A + \frac{1}{1 + \alpha} P''_{\lambda_n}(\btheta_\A) \right\} (\btheta_\A - \hat{\btheta}_\A)\right] + o(1)\\
&= \frac{1}{2} \tr\left[(\bsigma_\A^* + \bo{b}^{*} \bo{b}^{*T}) \left\{ \Psi_\A + \frac{1}{1 + \alpha} P''_{\lambda_n}(\btheta_\A) \right\}\right] + o(1), \label{taylor_1}
\end{align}
The final step is obtained from the asymptotic distribution of $\hat{\btheta}_\A$ using  Theorem \ref{theorem:asymp}. 
A Taylor series expansion of $n E[E[L_\alpha(\btheta |\X, \lambda_n) | \btheta = \hat{\btheta}_\A]]$ about $\btheta = \btheta_\A$  and using Equations (\ref{true_der}) and (\ref{2nd_der}), we get
\begin{align}
n  E[E[L_\alpha(\btheta &|\X, \lambda_n) | \btheta = \hat{\btheta}_\A]] - 	nE[L_\alpha(\btheta_\A |\X, \lambda_n)] \\
& = nE\left[(\hat{\btheta}_\A - \btheta_\A)E\left[\left[ \frac{\partial}{\partial \btheta} L_\alpha(\btheta |\X, \lambda_n) \right]_{\btheta = \btheta_\A}\right]\right] \\
& \ \ \ \ + \frac{n}{2}E\left[(\hat{\btheta}_\A - \btheta_\A)^T E\left[ \left[\frac{\partial^2}{\partial \btheta^T \partial \btheta} L_\alpha(\btheta |\X, \lambda_n) \right]_{\btheta = \btheta_\A} \right](\hat{\btheta}_\A - \btheta_\A)\right] + o(1)\\
&= \frac{n}{2}E\left[(\btheta_\A - \hat{\btheta}_\A)^T  \left\{ \Psi_\A + \frac{1}{1 + \alpha} P''_{\lambda_n}(\btheta_\A) \right\} (\btheta_\A - \hat{\btheta}_\A)\right] + o(1)\\
&=  \frac{1}{2}\tr\left[(\bsigma_\A^* + \bo{b}^{*} \bo{b}^{*T}) \left\{ \Psi_\A + \frac{1}{1 + \alpha} P''_{\lambda_n}(\btheta_\A) \right\}\right] + o(1). \label{taylor_2}
\end{align}
Substituting (\ref{taylor_1}) and (\ref{taylor_2}) in Equation (\ref{expr}), we find
\begin{align}
nE[E[D_\alpha(\btheta |\X, \lambda_n) | \btheta = \hat{\btheta}_\A]] =& nE[D_\alpha(\hat{\btheta}_\A |\X, \lambda_n) ] \nn
&+ \tr\left[(\bsigma_\A^* +  \bo{b}^{*} \bo{b}^{*T}) \left\{ \Psi_\A + \frac{1}{1 + \alpha} P''_{\lambda_n}(\btheta_\A) \right\}\right] + o(1).
\end{align}
Now, $E[D_\alpha(\hat{\btheta}_\A |\X, \lambda_n) ]$ is estimated by $D_\alpha(\hat{\btheta}_\A |\X, \lambda_n) $. From Equation (\ref{vi}), we observe that the first term in $V_i$ remains unchanged for all sub-models. Therefore, for $\alpha>0$, the robust AIC is expressed as
\be
RAIC = 
- \frac{1+\alpha}{ \alpha} \sum_{i=1}^n f_i^\alpha + \tr\left[(\hat{\bsigma}_\A^* +  \hat{\bo{b}}^{*} \hat{\bo{b}}^{*T}) \left\{ \hat{\Psi}_\A + \frac{1}{1 + \alpha} P''_{\lambda_n}(\hat{\btheta}_\A) \right\}\right],
\ee
where $\hat{\bsigma}_\A, \ \hat{\bo{b}}^*$, $\hat{\Psi}_\A$, $\hat{\xi}_{\alpha}$ and $\hat{\xi}_{2\alpha}$ are the estimates of $\bsigma_\A, \ \bo{b}^*$ and $\Psi_\A$, $\xi_{\alpha}$ and $\xi_{2\alpha}$ respectively. 
For $\alpha=0$, it becomes
\be
RAIC = 
-  \sum_{i=1}^n \log(f_i) + \tr\left[(\hat{\bsigma}_\A^* + \hat{\bo{b}}^{*} \hat{\bo{b}}^{*T}) \left\{ \hat{\Psi}_\A +  P''_{\lambda_n}(\hat{\btheta}_\A) \right\}\right].
\ee

\end{document}